\documentclass[preprint,12pt]{elsarticle}

\usepackage{lineno}

\usepackage{epsfig}
\usepackage{graphicx}
\usepackage{subcaption}
\usepackage{ragged2e}
\usepackage{amssymb}
\usepackage{fixltx2e}
\usepackage{textcomp}

\journal{Nuclear Physics B}

\begin{document}

\begin{frontmatter}



\title{Silicon dual pillar structure with a distributed Bragg reflector for dielectric laser accelerators: Design and fabrication}


\author[1]{Peyman Yousefi}
\author[1]{Joshua McNeur}
\author[1]{Martin Koz\'{a}k}
\author[2]{Uwe Niedermayer}
\author[3]{Florentina Gannott}
\author[3]{Olga Lohse}
\author[2]{Oliver Boine-Frankenheim}
\author[1,3]{Peter Hommelhoff}

\address[1]{Department of Physics, Friedrich Alexander University Erlangen Nuremberg, Staudtstr. 1, 91058 Erlangen, Germany}
\address[2]{TEMF, TU Darmstadt, Schlossgartenstr. 8, 64289 Darmstadt, Germany}
\address[3]{Max Planck Institute for the Science of Light, Staudtstr. 2, 91058 Erlangen, Germany}

\begin{abstract}
Dielectric laser accelerators (DLAs) have proven to be good candidates for miniaturized particle accelerators. They rely on micro-fabricated dielectrics which are able to modulate the kinetic energy of the incoming electron beam under a proper laser illumination. In this paper we demonstrate a dual pillar structure with a distributed Bragg reflector to mimic a double sided illumination to the electron path. The structure is fabricated by an electron beam lithography technique followed by a cryogenic reactive ion etching process. Such a structure can accelerate the injected 28 keV electrons by a gradient of approximately 150 MeV/m which can be further optimized towards the GeV/m regime.
\end{abstract}


\end{frontmatter}


\section{Introduction}
Dielectric laser acceleration is a novel concept to accelerate charged particles using short laser pulses in the presence of dielectrics structured on the micrometer scale \cite{Peralta:13,Breuer:13}. The basic idea of this concept is to excite a traveling evanescent wave in the close vicinity of a dielectric structure that interacts with an incoming electron beam in a way that the electron velocity is matched with the phase velocity of the propagating wave. This will accelerate or decelerate electrons depending on their injection phase with respect to the synchronous near-field \cite{England:14}.
\par
Dielectrics with high laser induced damage thresholds such as Al\textsubscript{2}O\textsubscript{3} and quartz are preferred candidates to achieve a high acceleration gradient, however micro-fabrication processes of such dielectrics have not been fully developed. Silicon however is a well known material whose micro-fabrication processes have been studied for decades \cite{Tachi88CryogenicEtch,Jansen96RIEoverview}. Even though its damage threshold fluence is as low as 0.19 J/cm$^{2}$ for a 1 ps, 800 nm, 600 Hz laser \cite{Soong:12Damage} it can still provide a good starting point in the DLA field.

\section{Design and fabrication}
Different designs for DLA structures from single grating to complex woodpile structure have been demonstrated \cite{Breuer:14Experiment,Leedle:15Grating,Olav:14BuriedGrating,Wu:14,Kozak:17Plane}. It has been already shown that bonding two single side gratings to make a double sided grating can help to achieve a more uniform field profile to confine the electron beam inside the acceleration channel \cite{Peralta:12Dualgrating}. However bonding two gratings adds one more step in the fabrication process which makes it difficult to control the grating alignment with the necessary sub micron precision. Dual pillar structure is in fact a two sided grating which is fabricated in one step. This reduces the fabrication effort and allows for much better control over alignment. Further, the design reduces the amount of dielectric material that the laser pulse needs to traverse and thereby reduces losses, dispersion and nonlinear distortion of the pulse. Such a structure has already been proposed and demonstrated \cite{Leedle:15DualPillars}.
\par
A dual laser drive is essential for DLA to enhance the acceleration gradient and reduce defocusing forces along the acceleration channel \cite{Breuer:14Theory}. However, this is difficult to realize experimentally. Nevertheless, adding a distributed Bragg reflector (DBR) acting as a mirror after the dual pillars achieves many of the same effects without requiring two phase controlled counter propagating laser pulses. The DBR mimics a dual laser drive which in theory can increase the acceleration efficiency to about $70\%$ of the incident field for low energy electrons with $\beta=0.32$ \citep{Niedermayer:17DualPillarSim}. In this work we add a DBR on one side of the structure and illuminate the pillars by infrared laser pulses. 
Illuminating the pillars from the side will excite a traveling wave along the channel in between the pillars where the electron beam is injected. The grating period must be matched with the wavelength of the laser to ensure the velocity matching condition is met. Since we are planning to inject a sub-relativistic electron beam, therefore the phase velocity has to be reduced which means the grating period has to be smaller than the optical wavelength:

\par
\centering
$$ \lambda_{p}=\beta\lambda $$
\par
\justifying
where $\lambda_{p}$ is the grating period, $\beta$ is the electron velocity divided by the speed of light and $\lambda$ is the optical wavelength.
\par
Figure 1 shows the geometry of the dual pillar structure with a DBR. The grating period is designed with respect to the initial kinetic energy of electrons and the optical wavelength which are 28 keV and 2 $\mu$m, respectively. The DBR spacing is only a quarter of the laser wavelength and the thickness of the silicon layers are $\lambda$/(4\textit{n}) in order to effectively illuminate the dual pillar structure from the other side, where \textit{n} is the refractive index of silicon. The aperture size, \textit{g}, is set to 200 nm to allow for a sufficient current of transmitted electrons through the channel.
\begin{figure}
	\centering
	\includegraphics[width=0.3\textwidth, angle=-90]{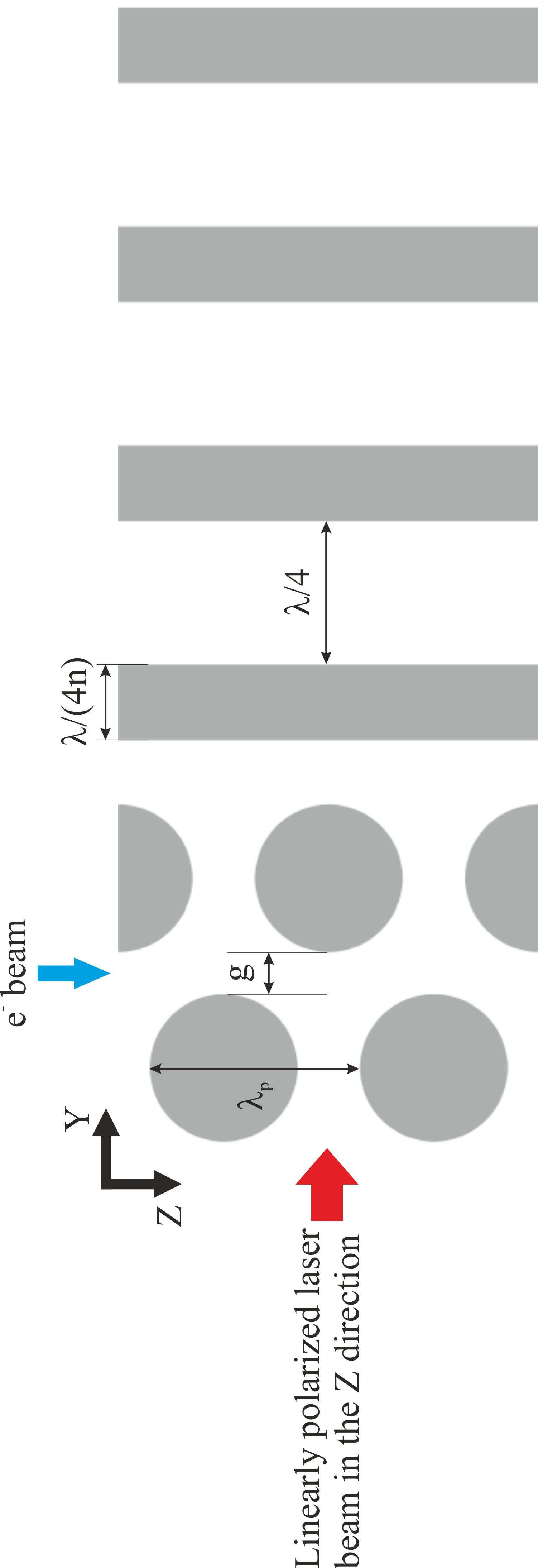}
	\caption[h]{The geometry of the dual pillar grating with a DBR. The grating period ($\lambda_{p}$) is given by the optical wavelength and electrons initial energy. The DBR spacing ($\lambda$/4) and the thickness of silicon layers ($\lambda$/(4\textit{n})) are given by the optical wavelength, where \textit{n} is the refractive index of silicon. The other distances to be optimized by simulation.}
\end{figure}
\par 
Electromagnetic simulations have been carried out by the Lumerical FDTD Solutions.
Figure 2a shows a simulation result of the longitudinal as well as the transverse components of the electric field profile after side illumination of the pillars by a 2 $\mu$m laser with a pulse duration of 100 fs. The DBR mimics a dual laser drive in the channel, reducing the percentage of electrons deflected as they travel through the channel.
\par
The fabrication process consists of two major steps namely, electron beam lithography and cryogenic reactive ion etching (RIE). We used a N-type silicon $<$100$>$ wafer doped with phosphorus. After a proper surface cleaning we spin coated a negative tone electron-beam sensitive resist and ran the electron beam exposure by a RAITH150 Two system. The resist thickness has to be thin enough for a good pattering resolution in the electron beam lithography step and thick enough for a proper etching durability in the deep silicon etching step. A directional etching was carried out after properly developing the resist. The cryogenic RIE process uses a gas combination of SF\textsubscript{6}:O\textsubscript{2} with a certain ratio at \mbox{-120\textcelsius{}} to directionally etch silicon \cite{Dussart14CryogenicEtch}. We etched the structure to a depth of ~3.0 $\pm$0.1 $\mu$m with an Oxford Instruments Plasmalab system 100. Fabrication precision and uniformity are important to have an optimal DLA structure. Non-uniformity for instance in the pillars diameters can affect the field profile which can lower the acceleration efficiency. To prevent that to occur a so called proximity effect correction method is applied to correct the exposure dose distribution in the electron beam lithography step to compensate the unwanted exposure due to the interaction of the primary beam with the resist and the substrate \cite{Chang:75Proximity}. Our final structures have a standard deviation of 10 nm from the designed geometry.
\par
\begin{figure}
	\centering
    \begin{subfigure}{0.45\textwidth}
        \includegraphics[width=\textwidth]{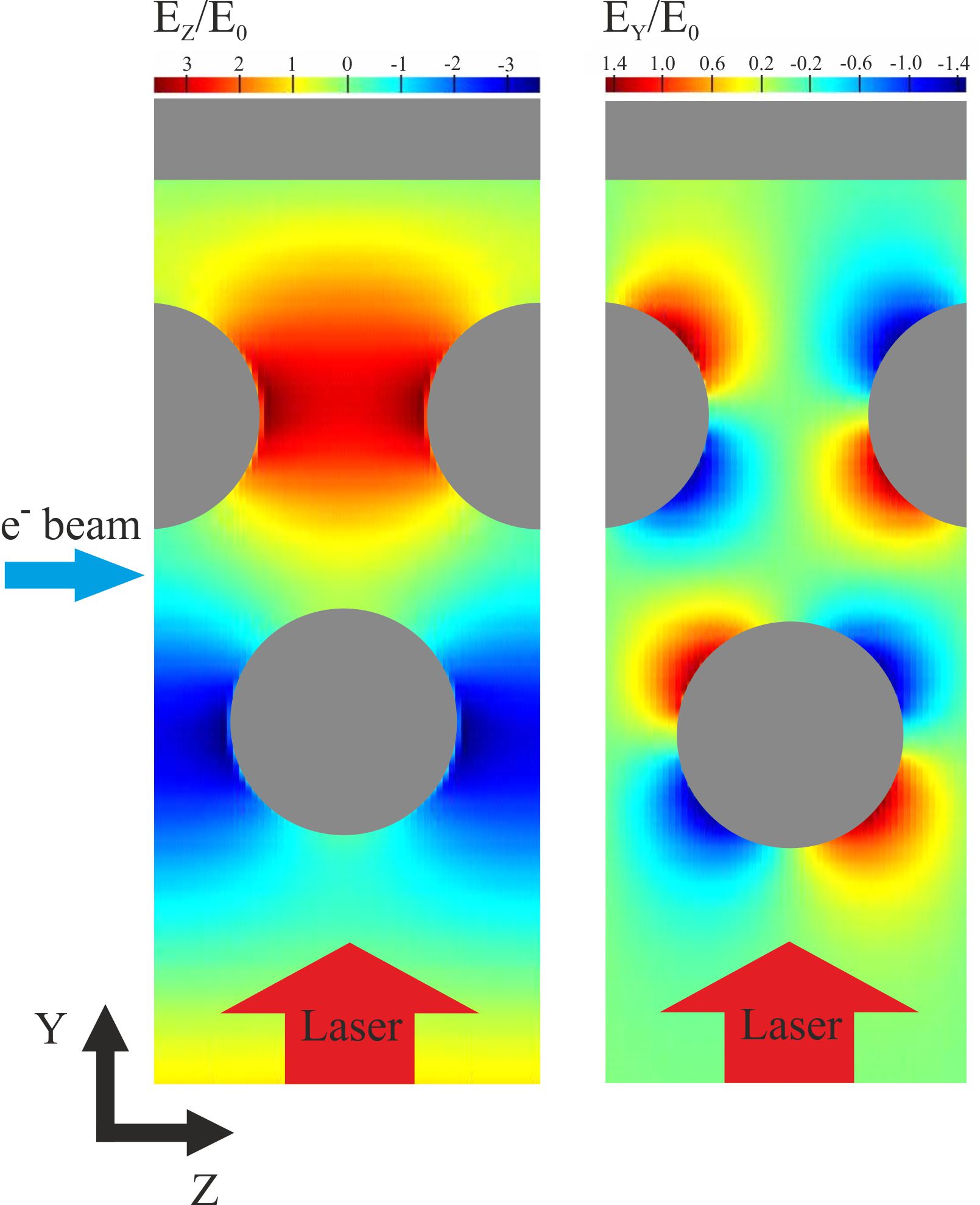}
       	\caption[H]{}
       
    \end{subfigure}\hspace{10mm}
    \begin{subfigure}{0.46\textwidth}
        \includegraphics[width=\textwidth]{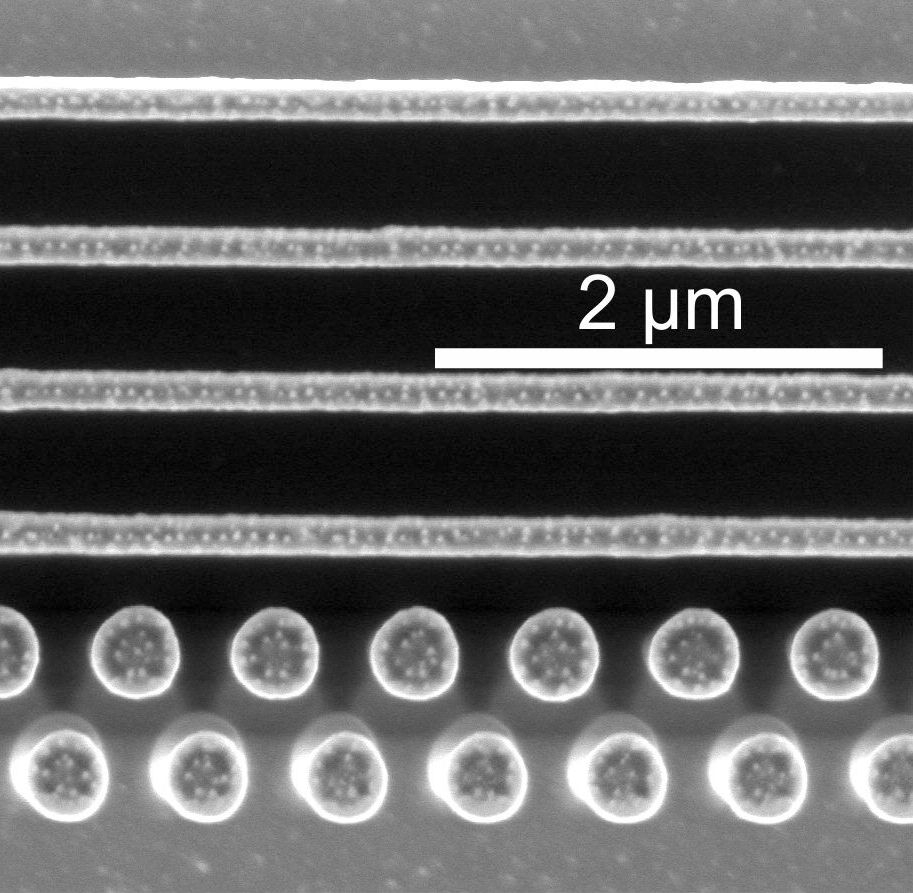}
       	\caption[H]{}
    \end{subfigure}
        \caption[H]{Dual pillar structure with DBR. Laser is illuminated from the side and electron beam is injected through the pillars. a) Simulation of the longitudinal (\textit{E$_{Z}$}) and the transverse (\textit{E$_{Y}$}) electric field for a 2$\mu$m laser with 100 fs pulse duration. b) Final structure out of silicon. The bright spots on top correspond to the resist residues after the etching step.}
\end{figure}
\par
Figure 2b illustrates the final structure fabricated out of silicon. It has a grating period of 620 nm with pillar diameter of 360 nm and a DBR consisting of four 130 nm thick layers of silicon spaced 500 nm apart. The resist has not been removed after the etching step which can be seen as bright spots in the figure. It can be removed by a proper oxygen plasma treatment, however that would grow a thin layer of an oxide layer on the whole structure which could change its geometry and possibly lead to slight charge accumulation during the experiment. It would be ideal to optimize a proper post-process cleaning for future fabrication attempts.

\par
\begin{figure}
	\centering
        \includegraphics[width=100mm]{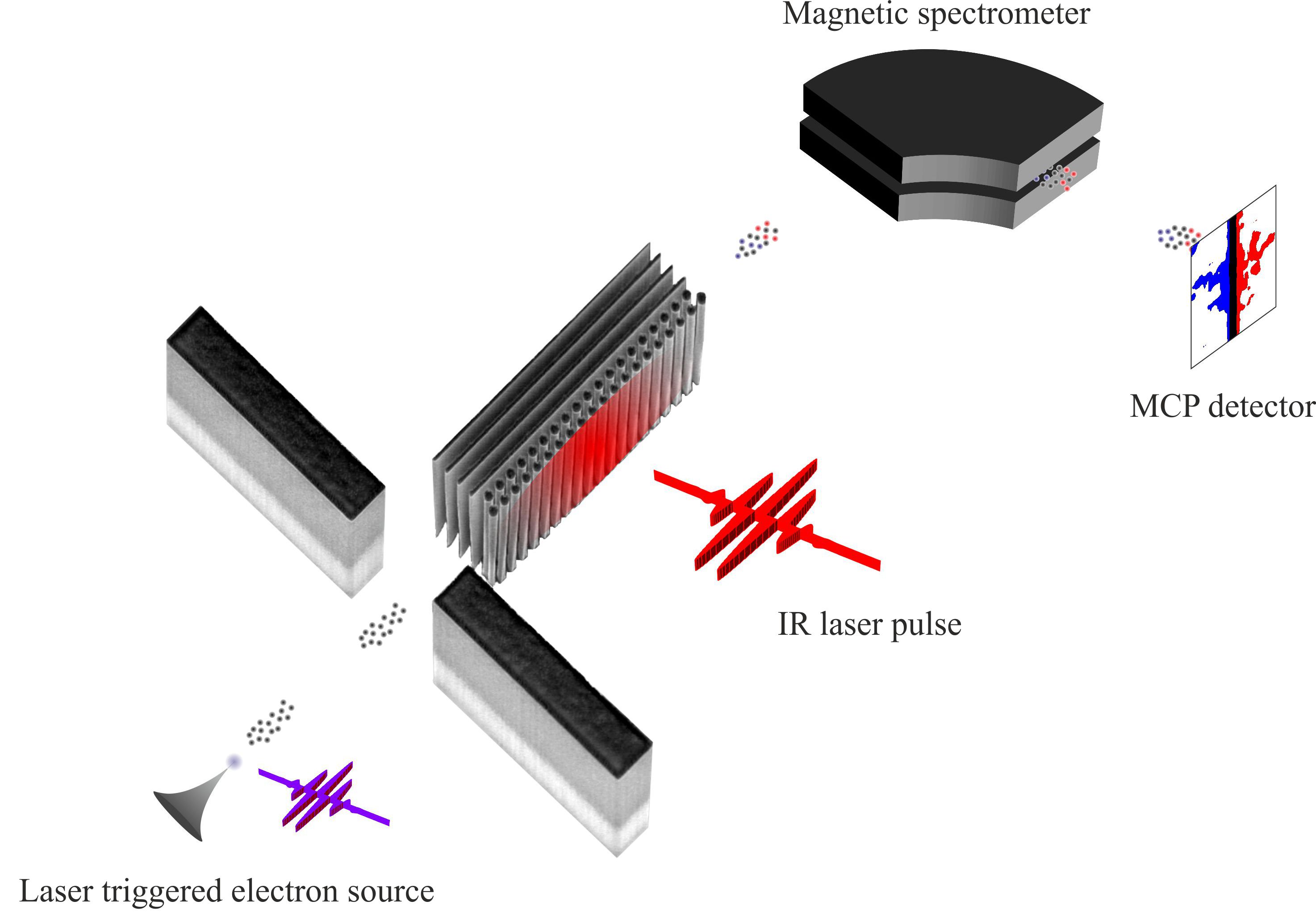}
       	\caption[h]{Schematic illustration of the experimental setup}
\end{figure}
\par
\section{Experimental setup}
The dual pillar structures are illuminated by laser pulses with a central wavelength of 1930 nm, a pulse duration of 100 fs and a repetition rate of 1 kHz and an intensity determined by the damage threshold of silicon. The electron source is a schottky SEM (Phillips xl50) whose schottky tip is illuminated by pulsed UV lasers, resulting in a pulsed electron beam. This 28 keV beam is focused to a spot size of 10 nm at the point of interaction with the dual pillar structure. Figure 3 shows our experimental setup to accelerate sub-relativistic electrons by silicon dual pillar structures with DBR. Electron bunches undergo an energy modulation after interacting with the incident electric field in between the pillars. We use a magnetic spectrometer and a micro-channel plate (MCP) detector for monitoring the energy modulation. The two alignment walls were placed in front of the structure for in situ alignment of the electron beam through the channel. A more detailed description of the experiment and its results will be reported elsewhere.

\par
Our simulations suggest an acceleration gradient of about 150 MeV/m which can be even further optimized towards GeV/m regime if the gratings were tapered with respect to the kinetic energy of the injected electrons. New geometries of dual pillars to accelerate electrons more efficiently are under active investigation. A complete accelerator on a chip however will need other components such as focusing elements to deliver a high quality beam as an output. Miniaturizing all the components on chip would make it possible to have a compact accelerator that can open up many new applications in different fields from low energy medical applications to high energy fundamental physics.

\section{Acknowledgment}
We thank the Max Planck institute for the science of light for providing us its clean room for all the micro-fabrication steps.
\par
This work was supported by the Gordon and Betty Moore Foundation under grant GBMF4744 (Accelerator on a Chip International Program).




\bibliographystyle{elsarticle-num} 
\bibliography{MyBib}





\end{document}